# Adhesive forces in droplet kinetic friction

Glen McHale[1], Sara Janahi[1], Hernán Barrio-Zhang[1], Yaofeng Wang[1], Jinju Chen[2],
Gary G. Wells[1] and Rodrigo Ledesma-Aguilar[1]

[1]*Institute for Multiscale Thermofluids, School of Engineering,*
*The University of Edinburgh, Edinburgh, EH9 3FB, UK.*
[2]*Department of Materials, Loughborough University, Loughborough, LE11 3TU, UK*

**Abstract**

Kinetic frictional forces resisting droplet motion often appear to be separate to surface wettability and adhesive forces. Here we show that such friction arises from a simple combination of the contact angle hysteresis and adhesive force. We show theoretically, and confirm using tilt angle experiments of droplets on liquid-like surfaces, the dependence of the coefficient of droplet-on-solid kinetic friction on system parameters. We also show that a molecular kinetic-type model can describe the observed non-linear velocity-force relationship. Our findings provide a fundamental understanding of the relationship between droplet-on-solid friction, and wettability and liquid adhesion.

**Article Text**

*Introduction* – From droplets of rain on a window to droplets of coffee on a table, small sessile droplets resting on surfaces are ever-present in society. Understanding these situations is underpinned by the studies of Thomas Young in 1805 which related the three interfacial tensions separating the solid, liquid and gas interfaces to the observable angle of contact at the three-phase contact line where the droplet meets the supporting solid [1]. However, if Young's Law held perfectly for practical surfaces, a small droplet of rain would always slide down a window and drying droplets would never leave ring stains [2]. Describing such problems arising from contact line pinning leads to the apparent abandonment of Young's theoretical equilibrium contact angle and the use of empirically observed advancing and receding contact angles [3]. These two angles describe the possible range of contact angles - the contact angle hysteresis - between which the measured static contact angle will be found in any deposition process. Contact angle hysteresis is a characteristic of any practical surface and represents the variation from a perfectly smooth and homogeneous surface due to roughness and surface chemistry. The force pinning a static droplet can be considered a static friction force with the resistance to droplet motion being the corresponding dynamic droplet-on-solid friction in analogy to the separation of static and kinetic regimes for one solid sliding on another [4]. However, unlike solids, droplets can be shaped whilst conserving volume and so it is possible to use the contact angle hysteresis to pre-stretch a droplet into its dynamic equilibrium shape so the maximum static force no longer exceeds the force for maintaining motion [5].

Recently, droplet-on-solid kinetic friction has been studied through measuring forces on cantilevers as a substrate is displaced and by motion of droplets on inclined planes [6–8]. It has been suggested that an empirical law relating the droplet-on-solid friction to the speed of droplet motion might apply over the parameter range examined and from those results a dimensionless friction coefficient, *β*, has been suggested as a material parameter [8]. Whilst practical surfaces always exhibit some heterogeneity, there has been significant recent progress in techniques to create smooth ultra-low contact angle hysteresis surfaces using covalently-attached liquid-like polymer chains [9–11]. It has also been shown that droplets of water on polydimethylsiloxane (PDMS) liquid-like surfaces can have similar static contact angles, but significantly different speeds of motion once static friction is overcome [12,13]. Such surfaces where kinetic friction can be converted from high to low, offer a unique opportunity to investigate the relationship between wettability, and droplet-on-solid kinetic friction and droplet adhesion. Here, we show a first-principles approach can be taken to defining coefficients of droplet-on-solid friction from the ratio of frictional and adhesive forces with excellent agreement with experimental observations on low and high droplet-on-solid kinetic friction surfaces. We also show the experimentally observed droplet kinetic friction-speed relationship for droplets of water on liquid-like surfaces is non-linear and can be described by a molecular kinetic theory (MKT) type model [14,15], which inherently encapsulates the droplet adhesion. The first order expansion of our MKT type model provides a linear approximation which incorporates the previous empirical law.

*Droplet-on-solid friction and the adhesive force* – For a solid object sliding on another solid the weight of the object due to gravity is translated into frictional forces opposing motion through static and kinetic coefficients of friction following Amontons' law [16,17]. However, for a droplet the size characterizing a volume of liquid is smaller than the capillary length $l_c=(\gamma_{LV}/\rho g)^{1/2}$, where $\gamma_{LV}$ and $\rho$ are the liquid-vapor interfacial tension and the density of the liquid, respectively and $g$ is the acceleration due to gravity, capillary forces dominate over the force due to gravity. We therefore define the coefficient of droplet friction, $\mu$, as the ratio of the frictional force, $F_f$, to the droplet adhesive force from the normal component, $F_n$, of the capillary force using an Amontons'-like law [18], i.e.

$$\mu \equiv \frac{F_f}{F_n} \tag{1}$$

where $\mu$ is either a static coefficient, $\mu_s$, or a kinetic coefficient, $\mu_k$, respectively, when the droplet is static or in motion. We now assume a small distortion from a spherical cap cross-sectional profile shape in the *x*-z plane [Fig. 1(a)] due to an applied force in the *x*-direction, and use a smooth and continuous parameterization of the contact angle, $\theta(\varphi)$, around a circular three-phase contact line from its maximum value at the front, $\theta(0)=\theta_f$, to its minimum value at the back, $\theta(\pi)=\theta_b$,

$$\cos\theta(\varphi) = \cos\theta_f + \frac{1-\cos\varphi}{2}(\cos\theta_b - \cos\theta_f) \tag{2}$$

The in-plane component of the solid-liquid interfacial tension force per unit length directed towards the centre of the droplet at a point on the contact line parameterized by $\varphi$ is $f_\gamma = \gamma_{LV}\cos\theta(\varphi)$ and the *x*-component is $f_x = \gamma_{LV}\cos\theta(\varphi)\cos\varphi$. Integrating this force around the contact line gives a Furmidge-type equation for the frictional force [19–21], as shown in the Supplemental Material S1,

$$F_f = 2kr_c\gamma_{LV}(\cos\theta_b - \cos\theta_f) \tag{3}$$

where the shape factor is $k=\pi/4\approx 0.785$ for a circular contact area. For a static droplet on the cusp of motion, the frictional force is the contact pinning force, and the front and back contact angles in Eq. (3) are equal to the static advancing and receding contact angles, $\theta_f=\theta_a$ and $\theta_b=\theta_r$, respectively. For a droplet in motion, the frictional force is given by the measured dynamic front and back contact angles, which are dependent on the speed of droplet motion. A similar evaluation of the normal forces around the circular three-phase contact line gives $F_n = 2\pi r_c\gamma_{LV}\sin\theta_{ave}$, where $\theta_{ave}=(\theta_f+\theta_b)/2$ is the average contact angle [Supplemental Material S1].

Physically, the shape factor, $k$, can be interpreted as a weighting factor for combining the net force in the $x$-$y$ plane in the direction of distortion arising from cross-sectional profiles taken from one side of the droplet to the other, each with contact angles $\theta_f = \theta(\varphi)$ and $\theta_b = \theta(\pi - \varphi)$) [Fig. 1(b)]. Importantly, our shape factor, $k$, does not represent an elongation or distortion of the contact area, which is often its interpretation in the literature relating to the Furmidge equation. However, if there is a known distortion of the three-phase contact line from circular, a shape factor could still be calculated; an example for an elliptical contact line is given in the Supplemental Material S1.

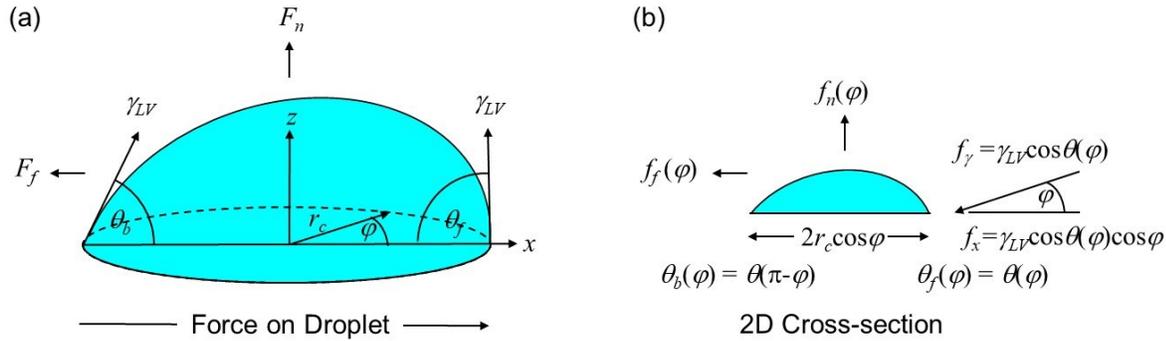

FIG. 1. Response of a droplet to an applied force. (a) Normal component of liquid-vapor interfacial tension force, $F_n$, and the in-plane frictional force, $F_f$, acting on a droplet distorted in the $x$-$z$ plane by an applied force. (b) Interpretation of forces on the three-phase contact line as arising from two-dimensional cross-sectional profiles each with contact length $r_c\cos\varphi$ and front and back contact angles $\theta(\varphi)$ and $\theta(\pi-\varphi)$, respectively, at positions parameterized by the angle $\varphi$ along the contact line.

From our results, the coefficient of droplet-on-solid friction, $\mu$, only depends on the two contact angles ($\theta_f$, $\theta_b$) and the shape factor, $k$. Alternatively, we can use the average contact angle $\theta_{ave}$, and the contact angle difference $\Delta\theta=(\theta_f -\theta_b)$, to expand the cosines in Eq. (3) to first order, so the coefficient of friction becomes,

$$\mu = \frac{k\Delta\theta}{\pi} \qquad (4)$$

In the context of Eq. (1), wettability defined through the average contact angle also defines the normal component of droplet capillary (adhesive) force and hence the adhesion of the droplet to the surface. The role of the coefficient of friction, Eq. (4), is to translate the liquid-solid adhesive force normal to the surface into friction in the plane of the surface. The preceding equations describing droplets can be applied whether a droplet is static or in motion and so can be used to define both a static coefficient of friction, $\mu_s$, or a kinetic coefficient of friction, $\mu_k$. The difference is whether the average contact angle is a static contact angle $\theta_s$, approximating the one expected from Young's law, i.e. $\cos\theta_e = (\gamma_{SV} - \gamma_{SL})/\gamma_{LV}$ [1,22], and the advancing, $\theta_a$, and receding, $\theta_r$, contact angles are assumed, or whether the average contact angle is a dynamic contact angle $\theta_d$ and the dynamic front $\theta_f$ and back $\theta_b$ contact angles are used. We would not expect the coefficient of droplet-on-solid kinetic friction defined through Eq. (1) to be a material parameter because the difference $\Delta\theta$ would, in general, depend on the speed of droplet motion. It can, however, be considered as a system parameter dependent on the interactions between the liquid and the surface, including surface heterogeneity.

From the perspective of surface design, a low droplet-on-solid static friction surface can be achieved either by reducing the normal component of the capillary force or by reducing the contact angle hysteresis. Thus, both low and high droplet adhesion surfaces can be created with low droplet-on-solid static friction.

This idea can be shown explicitly by using the spherical cap volume, $\Omega$, to relate the contact radius to the contact angle and replacing the contact angle by the work of adhesion, $W_{adh} = \gamma_{LV}(1 + \cos\theta_e)$ in the normal component of the capillary force, i.e.

$$F_n = 2\pi\gamma_{LV} \left(\frac{3\Omega}{\pi\left(1+\frac{W_{adh}}{\gamma_{LV}}\right)}\right)^{1/3} \left(\frac{W_{adh}}{\gamma_{LV}}\right)\left(2 - \frac{W_{adh}}{\gamma_{LV}}\right)^{1/3} \quad (5)$$

In the limit of a perfectly superhydrophobic surface with $\theta_e \rightarrow 180°$, the work of adhesion and the normal component of the capillary force both vanish. Similarly, in the limit of a perfectly hydrophilic surface with $\theta_e \rightarrow 0°$, the work of adhesion and the normal component of the capillary force also vanish. Hence, in both these limiting cases, including the perfectly hydrophilic one corresponding to a film-forming surface, the droplet-on-solid friction is zero irrespective of the contact angle hysteresis, $\Delta\theta_{CAH}=\theta_a - \theta_r$, arising from surface heterogeneity. The maximum in the normal adhesive force for a spherical cap droplet of fixed volume occurs for a hydrophilic surface with a contact angle $\theta_e \approx 65.5°$ corresponding to a work of adhesion $W_{adh}=2^{1/2}\gamma_{LV}$. The maximum occurs at a contact angle below 90° because of the dependence of the contact radius, $r_c$, on the contact angle for a fixed volume. For a surface slippery to droplets, minimizing the contact angle hysteresis (or, equivalently, the coefficient of droplet-on-substrate static friction) is critical to minimizing the coupling of the adhesive force into the droplet-on-solid frictional force.

*Experiments on Liquid-like Surfaces* – To explore whether the coefficient of kinetic friction defined by Eq. (1) is proportional to the difference between the dynamic front and back contact angles [Eq. (4)], we created a set hydrophobic liquid-like surfaces on glass substrates and conducted tilt angle experiments [Fig. 2(a)] [Supplemental Material S2]. These surfaces were created using a "grafting-from" acid-catalyzed polycondensation procedure of dimethyldimethoxysilane to create a layer of covalently-attached hydroxy-terminated polydimethylsiloxane (PDMS) chains [Fig. 2(b)], i.e. a slippery omniphobic covalently-attached liquid-like surface (SOCAL), with ultra-low contact angle hysteresis $\Delta\theta_{CAH} \sim 2°$ [23]. Although these surfaces have ultra-low contact angle hysteresis (droplet-on-solid static friction), once droplets are set in motion they move slowly and so demonstrate high droplet-on-solid kinetic friction [12]. This has previously been attributed to the interaction between water and the silanol groups in the hydroxy-terminated PDMS chains [13]. We therefore used a chlorotrimethylsilane molecular-capping (methylation) procedure, which is believed to convert the silanol-termination of the PDMS chains to trimethyl-terminations (c-SOCAL), to create a set of additional surfaces with lower kinetic friction [Fig. 2(c)] [13]. The effectiveness of this vapor-phase capping procedure is variable depending on the process parameters, particularly the relative humidity (RH). We created surfaces using RH=30%, and 40% and 50% (denoted by c30-SOCAL, c40-SOCAL and c50-SOCAL) and found a significant reduction of the kinetic friction for the c50-SOCAL surface. The capping process makes almost no difference to the measured hydrophobicity of the samples for which the static contact angle increases from ~103° to ~106°, although the static contact angle hysteresis increases from ~2° to ~5° on the c50-SOCAL surface, which we interpret as due to the effect of the HCl by-product of the capping process on the surface homogeneity.

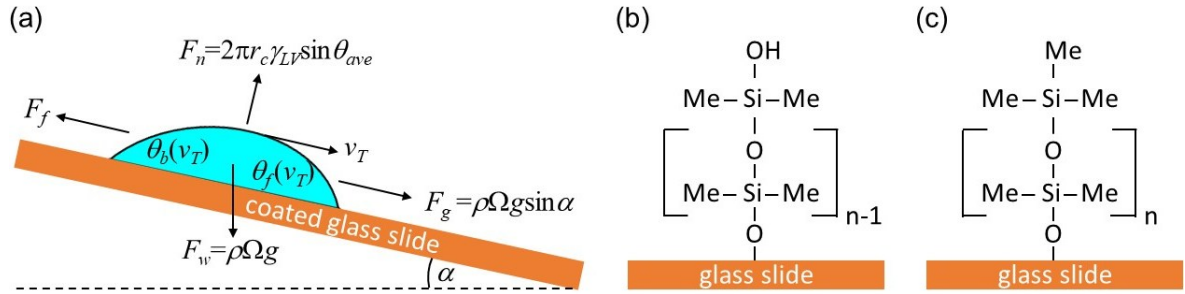

FIG 2. Measurements of droplet kinetic friction on hydrophobic PDMS-based liquid-like surfaces. (a) Tilt angle experiments with friction and normal adhesive forces measured under constant speed motion. (b) Ultra-low hysteresis covalently-attached OH-terminated PDMS chains providing a coating with high droplet kinetic friction. (c) Methyl-terminated PDMS chains obtained by using a molecular capping procedure to convert the surface in (b) into a low droplet kinetic friction surface.

We then recorded the motion of water droplets on surfaces tilted from the horizontal at a range of angles, $\alpha$, up to 40° and measured the front and back dynamic contact angles, and the steady state speed, $v_T$, at which the frictional force, $F_f$, balances the component of gravitational force along the surface, $F_g=\rho\Omega g\sin\alpha$. Figure 3 (a) shows data for the observed speed of 20 µl droplets on the liquid-like surfaces for a range of tilt angles up $\alpha$=40° corresponding to frictional forces, $F_f=F_g$, of around 126 µN; the kinetic frictional force is a non-linear function of the speed. For droplets on the high (SOCAL) and low (c50-SOCAL) kinetic friction surfaces tilted at 40°, the speeds are $v_T$=0.58 mm s$^{-1}$ to $v_T$=35 mm s$^{-1}$, respectively. This is a remarkable 60-fold increase induced by the capping procedure despite the increase in the static contact angle hysteresis (i.e. droplet-on-solid static friction). We did not observe any significant deviation from a circular contact area shape. To explore whether Eq. (4) relating the coefficient of droplet-on-solid kinetic friction, $\mu_k$, to the difference in dynamic front and back contact angles, $\Delta\theta$, is valid experimentally, we calculated the normal component of the surface tension force, $F_n$, from the average of the front and back contact angles and the observed contact radius of the droplet. In Fig. 3(b) we plot the coefficient of droplet-on-solid kinetic friction, $\mu_k$, determined by the ratio of forces in Eq. (1), and $\Delta\theta/\pi$ using the measured difference in front and back contact angles over a range of tilt angles up to $\alpha$=40°. The data follows a linear relationship with an experimentally determined slope of $k$=0.79±0.04, which is in excellent agreement with the predicted value of $k=\pi/4$=0.785 from Eq. (3). Prior values derived in the literature for the shape-factor $k$ in the Furmidge equation lie between $2/\pi$=0.637 [24] and $24/\pi^3$=0.774 [25]. However, these have assumed either non-continuous contact distributions along the contact line and/or an elongated droplet contact line shape of some form. In our approach, the shape factor $k$ is a factor allowing conversion from a two-dimensional model profile as shown in Fig. 1(b) to the three-dimensional droplet shape as shown in Fig. 1(a). Any elongation or distortion of the contact area shape from circular will result in a different value for $k$ but, providing reflection symmetry is retained in the *x-z* axis, Eq. (4) will remain valid.

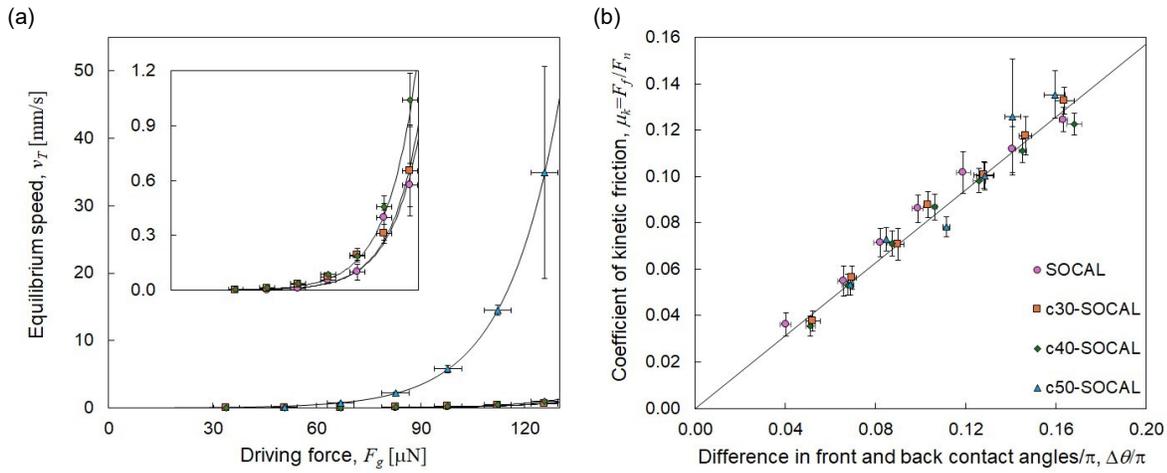

FIG. 3 (a) Equilibrium speeds of motion of 20 µl droplets on high (SOCAL, c30-SOCAL and c40-SOCAL) and low (c50-SOCAL) droplet kinetic friction liquid-like surfaces at substrate tilt angles up to 40° to the horizontal; solid lines are fits of the speed to a molecular-kinetic theory type model (Eq. (6)). (b) Linearity of the coefficient of droplet-on-solid kinetic friction, $\mu_k$, determined using measurements of $r_c$, $\theta_f$ and $\theta_b$; solid line is the predicted slope of $k=\pi/4$.

We now consider the relationship between the dynamic front and back contact angles, $\theta_f$ and $\theta_b$, and the droplet speed, $v_T$. Both contact angles change with speed, but the back contact angle is more sensitive than the front contact angle to the speed [Fig. 4(a)]. The back contact angle for droplets moving on the low kinetic friction c50-SOCAL surface has a much smaller change than for the other higher kinetic friction surfaces. However, when the data is considered using the coefficient of droplet-on-solid kinetic friction, $\mu_k$, determined by the ratio of forces in Eq. (1), and plotted using a logarithmic axis for the droplet speed, $v_T$, the shape of the observed curves on all surfaces is similar [Fig. 4(b)]; normalizing the velocities on each surface by a factor, $v_s$, allows the data to be overlaid when plotted using $\log_{10}(v_T/v_s)$ [Fig. 4(b) Inset]. The same conclusions can be reached by calculating the coefficient of droplet-on-solid kinetic friction using the measured difference in the front and back contact angles $\Delta\theta$. Thus, in these tilt angle experiments, similar values of coefficients of droplet-on-solid friction are achieved by droplets of the same volume on different liquid-like surfaces tilted at the same angle, but to do so the droplet has different terminal speeds and, hence, different front and back contact angles.

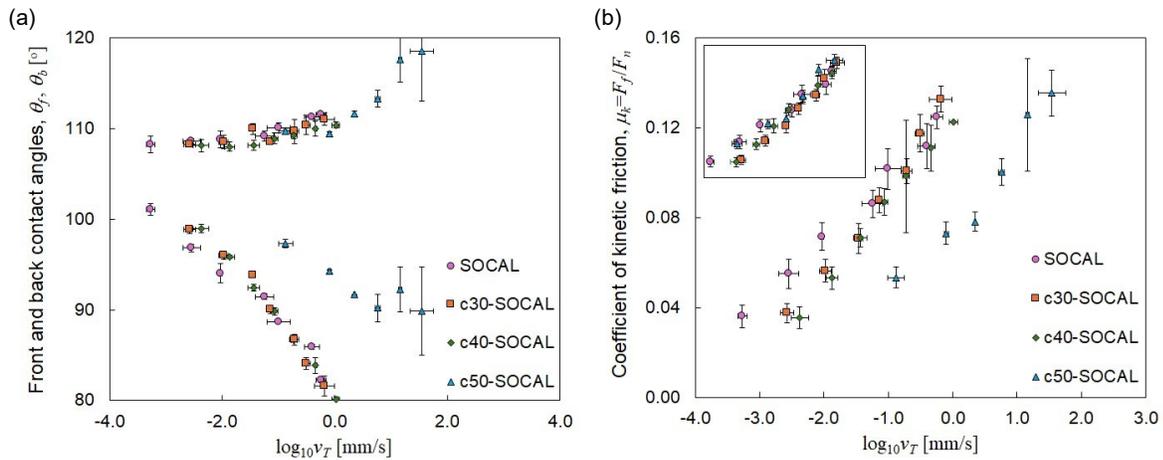

FIG. 4 (a) Velocity dependence of dynamic front and back contact angles for 20 μl droplets on high (SOCAL, c30-SOCAL and c40-SOCAL) and low (c50-SOCAL) droplet kinetic friction liquid-like surfaces at substrate tilt angles up to 40° to the horizontal. (b) Velocity dependence of coefficient of droplet-on-solid kinetic friction; Inset uses a $\log_{10}(v_T/v_s)$ axis, where $v_s$ are the parameters from the molecular kinetic theory (MKT) model, to overlay the data.

*A molecular-kinetic theory (MKT) type model* – It has previously been suggested molecular-kinetic theory (MKT) may apply to the motion of contact lines on liquid-like surfaces [14,15]. To understand the non-linear response of the droplets to the unbalanced capillary force driving the motion and the droplet-on-solid friction, we now develop a modified MKT-type model formulated in terms of an activation energy, $W(\varphi)$, which varies around the three-phase contact line (Fig. 5). At the three-phase contact line the molecules of water must be in a dynamic equilibrium with both forward and backward displacements (jumps). Similar to the original MKT model [14,15], the average distance of a molecular displacement of the contact line corresponds to the average distance between adsorption sites on the surface, $\lambda$, and the displacement occurs with an average frequency $K_o$. However, we differ from the original MKT model in defining the activation energy by using the dynamic equilibrium contact angle and a local contact angle, parameterized by Eq. (2) to capture the variation around the contact line, bounded between the dynamic front and back contact angles, rather than using the dynamic contact angle and the static contact angle,

$$W(\varphi) = \lambda^2 \gamma_{LV}(\cos\theta_d - \cos\theta(\varphi)) \quad (6)$$

The advantage of this definition of the energy barrier is that it necessarily incorporates the friction-adhesion concept for the motion of the three-phase contact line because for small differences in contact angles, $W(\varphi) \approx \lambda^2 \gamma_{LV} \Delta\theta(\varphi) \sin\theta_d$, where $\Delta\theta(\varphi) = (\theta(\varphi) - \theta_d)$. Using the normal component of the capillary force this can be written as $w(\varphi) \approx \lambda^2 \mu(\varphi) f_n$, where $\mu(\varphi) = \Delta\theta(\varphi)$ is the local coefficient of droplet-on-solid friction for forward and backward displacements at the contact line position parameterized by $\varphi$. The difference between forward and backward jumps around the contact line then gives the net speed of motion of the contact line,

$$v_T = \frac{1}{2\pi r_c} \oint 2\lambda K_0 \sinh\left[\frac{W(\varphi)}{k_B T}\right] \cos\varphi \, d\phi \quad (7)$$

which can be written [Supplemental Material S3],

$$v_T = v_s \sinh\left[\frac{\lambda^2 F_g}{4\pi r_c k_B T}\right] \quad F_g > F_s \quad (8)$$

where $v_s = \lambda K_o$, $k_B$ is the Boltzmann constant and $T$ is the temperature. We fit the data in Fig. 3(a) using a single value of $\lambda$=1.33 nm for all data sets and separate values for the $v_s$= 0.37, 0.32, 0.68 and 20.5 μm/s for the SOCAL, c30-SOCAL, c40-SOCAL and c50-SOCAL surfaces, respectively. These fits correspond to molecular displacements occurring with average frequencies $K_o$=282, 242, 513 and 15442 Hz. The characteristic distance for displacement, $\lambda$, is in the nanometer range which is physically plausible. In this model, the impact of the molecular capping process on droplet-on-solid friction arises through the change in average frequency, which results in a ca. 55-fold increase in the pre-factor $v_s$. A possible physical interpretation is that the silanol groups on OH-terminated PDMS act as trapping sites, which retain molecules for longer periods of time than other groups and so decrease the average molecular jump frequency.

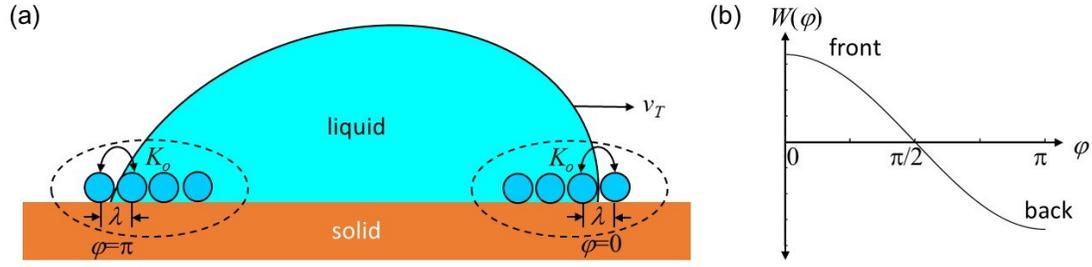

FIG 5. Schematic showing the molecular-kinetic theory type picture for the displacement of molecules of liquid at the three-phase contact line of the droplet. (a) Motion of molecules at the microscopic level between adsorption sites with average separation of $\lambda$ at an average frequency of $K_o$. (b) The energy barrier for forward and backward motion of molecules is proportional to the local coefficient of friction and the normal component of the liquid-vapor interfacial force.

We now consider the consistency between our coefficient of friction and an alternative empirical definition of a friction coefficient, $\beta$, given by $F_f = F_o + 2\beta r_c \eta v_T$, where $F_o$ is the extrapolation to zero speed, which has been suggested by Li *et al.* [8]. Clearly, the linear relationship between the friction force and the droplet speed suggested by this expression is not an accurate description of our data in Fig. 3(a). However, using a first order expansion our Eq. (8) gives a linear relationship and an equivalence of the models in this limit occurs by defining the friction coefficient as,

$$\beta = \frac{\pi k_B T}{2\eta K_o \lambda^3} \quad (9)$$

This suggests the pre-factor in Eq. (8) is given by $v_s = \pi k_B T / 2\eta \beta \lambda^2$ and so scales with the inverse of viscosity. This is also expected in the original MKT theory by using the Eyring theory for viscosity [26]. Assuming a $v_s \propto 1/\eta$ relationship also means a characteristic speed for liquid motion, $v^* = \gamma_{LV}/\eta$, arises naturally in the first order expansion of Eq. (8) from the surface tension dependence of the energy barrier.

*Conclusions* - In summary, our approach shows how droplet-on-solid kinetic friction can be related directly to the droplet adhesive force arising from wettability of the substrate surface. It provides an understanding of the shape factor in the frictional force as a translation from the forces in a two-dimensional cross-sectional droplet profile to the three-dimensional droplet system valid whether a droplet is static or in motion. Our experimental observations use ultra-low contact angle hysteresis liquid-like surfaces which appear almost identical from their static wettability, but which have speeds of motion which differ by more than an order of magnitude for the same driving force. From the observed force-velocity relationships, we have constructed a molecular-kinetic theory type model which incorporates the droplet-on-solid friction and through the coefficients of kinetic friction also includes the droplet adhesive force. Finally, the concepts presented are not limited to droplets, but are relevant to the motion of any three-phase contact line and to three-phase systems, such as flow in narrow capillaries and the motion of bubbles on surfaces.

*Acknowledgements* – We thank the UK EPSRC for financial support under grants EP/V049348/1, EP/V049615/1 and EP/V049615/2. YW acknowledges PhD studentship funding from the UK EPSRC (Project 2924258).

## Supplemental Material
## Adhesive forces in droplet kinetic friction

Glen McHale[1], Sara Janahi[1], Hernán Barrio-Zhang[1], Yaofeng Wang[1], Jinju Chen[2],
Gary G. Wells[1] and Rodrigo Ledesma-Aguilar[1]

[1]*Institute for Multiscale Thermofluids, School of Engineering,
The University of Edinburgh, Edinburgh, EH9 3FB, UK.*
[2]*Department of Materials, Loughborough University, Loughborough, LE11 3TU, UK*

### 1. Forces from Parameterization of the Contact Angle around a Circular or Elliptical Contact Line

*In-Plane Friction* - In the coordinate system of FIG. S1, the liquid-vapor interfacial (surface) tension force, and hence frictional force, in the *x-y* plane along the *x*-direction is,

$$F_f = -\gamma_{LV} \int_0^{2\pi} r(\varphi) \cos\theta(\varphi) \cos\varphi \, d\varphi \qquad \text{(S1.1)}$$

where $\varphi=0$ to $2\pi$ is the angle parameterizing the contact line, and $\theta_f=\theta(0)$ and $\theta_b=\theta(\pi)$ are the front and back contact angles.

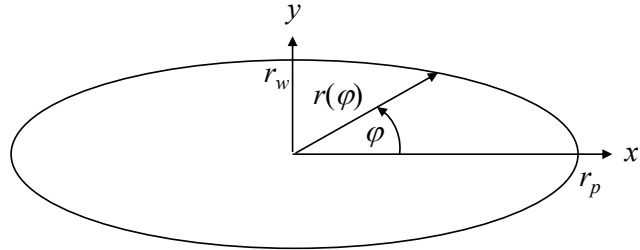

The simplest continuous and smooth parameterization is,

$$\cos\theta(\varphi) = \cos\theta_f + \frac{1-\cos\varphi}{2}\left(\cos\theta_b - \cos\theta_f\right) \qquad \text{(S1.2)}$$

and so,

$$F_f = -2\gamma_{LV} \int_0^{\pi} r(\varphi)\left[\cos\theta_f + \frac{1-\cos\varphi}{2}\left(\cos\theta_b - \cos\theta_f\right)\right]\cos\varphi \, d\varphi \qquad \text{(S1.3)}$$

Using the coordinate system $(x, y) = (r_p \cos\varphi, r_w \sin\varphi)$, the ellipse is given by,

$$\frac{x^2}{r_p^2} + \frac{y^2}{r_w^2} = 1 \qquad \text{(S1.4)}$$

where $r_p$ is the radius (half-length) seen in side profile viewing along the *y*-axis and $r_w$ is the radius (half-length) seen in side profile viewing along the *y*-axis. The eccentricity of the ellipse, *e*, is then given by,

$$e^2 = 1 - \frac{r_w^2}{r_p^2} \tag{S1.5}$$

The force along the major axis of the ellipse is then,

$$F_f = -2\gamma_{LV} \int_0^\pi r_p [\cos^2 \varphi + (1-e^2) \sin^2 \varphi]^{1/2} \left[ \cos \theta_f + \frac{1-\cos \varphi}{2}(\cos \theta_b - \cos \theta_f) \right] \cos \varphi \, d\varphi \tag{S1.6}$$

The first term vanishes and the integral reduces to,

$$F_f = -2\gamma_{LV} r_p (\cos \theta_b - \cos \theta_f) \int_0^\pi [\cos^2 \varphi + (1-e^2) \sin^2 \varphi]^{1/2} \left( \frac{1-\cos \varphi}{2} \right) \cos \varphi \, d\varphi \tag{S1.7}$$

When $r_w=r_p$ and the ellipse reduces to a circle with $e=0$, the integral evaluates to,

$$F_{f\_circle} = 2k_{circle}\gamma_{LV} r_p (\cos \theta_b - \cos \theta_f) \tag{S1.8}$$

where we have defined $k_{circle}=\pi/4\approx 0.785$ as the shape factor.

To evaluate the ellipse case, we define Eq. (S1.7) as,

$$F_f = 2k_{circle} I(e) \gamma_{LV} r_p (\cos \theta_b - \cos \theta_f) \tag{S1.9}$$

where the integral function $I(e)$ is defined as,

$$I(e) = -\left(\frac{4}{\pi}\right) \int_0^\pi [\cos^2 \varphi + (1-e^2) \sin^2 \varphi]^{1/2} \left( \frac{1-\cos \varphi}{2} \right) \cos \varphi \, d\varphi \tag{S1.10}$$

Thus, the shape factor for an elliptical contact area $k_{ellipse}=k_{circle} I(e)$ and can be calculated by a numerical evaluation of Eq. (S1.10). The expansion of the integral function about $e=0$ is,

$$I(e) \approx 1 + \frac{e^2}{8} - \frac{e^4}{64} + \frac{5e^6}{1024} - \frac{500e^8}{234057} + \tag{S1.11}$$

When the ellipse has $r_w>r_p$ and the ellipse is "fat" rather than elongated, the $e^2$ in Eq. (S1.10) and Eq. (S1.11) becomes negative.

*Normal Component of Surface Tension* - The normal component of the surface tension force in the $z$-direction is,

$$F_n = \gamma_{LV} \int_0^{2\pi} r(\varphi) \sin \theta(\varphi) \, d\varphi \tag{S1.12}$$

Taking first order terms in $(\theta_f-\theta_b)$, gives,

$$F_n \approx \gamma_{LV} \sin \theta_{ave} \, l_p(e) \tag{S1.13}$$

where $\theta_{ave}=(\theta_f+\theta_b)/2$ is the average of the front and back contact angles and the $l_p(e)$ is the perimeter (contact line) length defined by,

$$l_p(e) = r_p \int_0^{2\pi} [\cos^2 \varphi + (1-e^2) \sin^2 \varphi]^{1/2} d\varphi \tag{S1.14}$$

Writing in terms of the perimeter length of a circle,

$$F_n \approx 2\pi r_p \gamma_{LV} \sin \theta_{ave}\, l_{ratio}(e) \tag{S1.15}$$

where $\theta_{ave}=(\theta_f+\theta_b)/2$ is the average of the front and back contact angles and the $l_p(e)$ is the perimeter (contact line) length defined by,

$$l_{ratio}(e) = \left(\frac{1}{2\pi}\right) \int_0^{2\pi} [\cos^2 \varphi + (1 - e^2) \sin^2 \varphi]^{1/2} d\varphi \tag{S1.16}$$

This involves a complete elliptic integral of the second kind and can be evaluated numerically or through the approximation,

$$l_{ratio}(e) \approx 1 - \frac{e^2}{4} - \frac{3e^4}{64} - \frac{5e^6}{256} - \frac{175 e^8}{16384} - \tag{S1.17}$$

An alternative more accurate approximation is to use Ramanujan's approximation for the perimeter length of an ellipse [R1],

$$l_{ratio}(e) = \left(\frac{1}{2}\right)\left[3\left(1 + \frac{r_w}{r_p}\right) - \sqrt{\left(3 + \frac{r_w}{r_p}\right)\left(1 + \frac{3r_w}{r_p}\right)}\right] \tag{S1.18}$$

## 2. Materials and Methods

*Preparation of SOCAL Coated Glass Substrates* - SOCAL surfaces were created on 15 × 25 mm glass slides following the methodology detailed by Wang and McCarthy [R2] as optimized by Armstrong *et al.* [R3]. The glass slides were first cleaned in an ultrasonic bath using the following sequential process: (i) 10 minutes in a solution of 50 mL deionised (DI) water and 1mL Decon 90, (ii) 10 minutes in 30 mL DI water, (iii) 5 minutes in 30 mL acetone and (iv) 5 minutes in 30 mL isopropanol (IPA). The clean slides were then dried with compressed air and placed in an oxygen plasma oven (Henniker HPT-200) operating at 60 W for 20 minutes, adding hydroxyl (OH) groups to the glass substrate [R3]. The slides were manually dipped into a reactive solution of IPA, dimethyldimethoxysilane, and sulfuric acid (100, 10, and 1 wt% respectively) for 10 seconds and slowly withdrawn. The coated slides were then placed in a humidity-controlled environment at 60% ± 5% relative humidity (RH) and room temperature (T = 20–25°C) for 20 minutes, where the acid-catalysed graft polycondensation takes place, induced by the presence of the OH groups. The SOCAL surfaces were finally rinsed with DI water, IPA, and toluene to stop the reaction and remove excess unreacted material. This process provides a uniform ca. 4-5 nm thick coating of covalently-attached OH-terminated PDMS chains, which are non-cross-linked and retain chain flexibility, which is believed to cause the observed ultra-low contact angle hysteresis (typically ~1°-2°). Droplets of water moving on these surfaces appear to have high kinetic friction despite the ultra-low contact angle hysteresis [R4].

*Molecular-capping (Methylation) of SOCAL* – For the c30-SOCAL, c40-SOCAL and c50-SOCAL surfaces, the polar silanol groups in the SOCAL PDMS chains were capped with methyl groups following Khatir *et al.*'s [R5] procedure. The surfaces were first rinsed with DI water, IPA, and toluene and dried with compressed air to remove dust particles and precipitated salts. The SOCAL surfaces were placed in a bespoke RH chamber, where the airflow was adjusted to control the RH to 30%, 40% or 50%. 100 µL of chlorotrimethylsilane (≥ 99%) was poured onto a watch glass and the RH chamber was sealed for 2 hours, during which the single-step vapor deposition occurred at room temperature. The capped surfaces were then rinsed with DI, IPA, and toluene to wash away residuals. The relative humidity during the 2-hour procedure was not continuously monitored as preliminary tests revealed damage to the sensor caused by chlorotrimethylsilane. Therefore, the RH was monitored only initially when regulating the airflow to the chamber. However, further tests were conducted where the RH was monitored every 20 minutes to assess the error, indicating a 2% variation in the RH over the 2-hour interval.

*Contact Angle Measurements* - The liquid dispensing system consisted of a thin needle, 0.4 mm in outer diameter, connected to a 500 µL syringe (Hamilton) and a micropump (Cellix ExiGo) programmed through the SmartFlo software. To ensure no contaminants were present in the syringe, 10 mL acetone, IPA, and DI was injected through the needle prior to use. A sample was placed on a motorized tilt stage (Thorlabs, K10CR2/M), where the inclination angle was controlled using a Kinesis software interface (Thorlabs). A video camera (iDS UI-3160CP), operating at 169 frames per second (fps), was used to record a side view of the droplet and is connected to a computer for image acquisition and analysis. The advancing and receding contact angles of droplets were measured through volume addition and withdrawal experiments of ultrapure water at a fixed flow rate. An initial volume of $\Omega = 8$ µL of ultrapure water (Ultrapur Water, Supelco) was dispensed on the sample, followed by a 20 second rest period to adjust the needle; the needle was placed on one side of the droplet and the contact angle measurements are taken from the opposite edge, where a spherical cap shape is maintained. The camera recorded the addition of $\Omega = 4$ µL and a 30 second rest period, where the contact line reverts to a static position. Subsequently, a volume of $\Omega = 4$ µL was withdrawn and the droplet of ultrapure droplet was given another 30 second rest period, after which the video recording ends [R4]. This process was repeated three times at different positions on each SOCAL and capped SOCAL surface. The videos were processed through the pyDSA droplet shape analyzer software [R6].

To assess the kinetics of ultrapure water on SOCAL and capped SOCAL surfaces, droplet motion was induced by gravitational force. The inclination angle, $\alpha$, of the platform was first adjusted through the Kinesis software within the range of 5-40°. This range was chosen to determine the sliding angle of the ultrapure water droplets on the SOCAL and capped SOCAL surfaces, ensuring sufficient variation in droplet behavior was observed. The samples were thoroughly rinsed with DI water prior to performing the tilting experiments and dried with compressed air between each change in the inclination angle. This step was crucial to ensure the removal of dust particles and prevent residual water from artificially reducing the friction of the surface; a reduction in friction would have led to falsely increased droplet velocities across the different surface types during the trials. The tilting procedure was as follows. A volume of $\Omega = 20$ µL of ultrapure water was dispensed onto the tilted sample, with the needle serving as an anchor to hold it in place. The video recording was then started, and the needle slowly raised to release the droplet and allow it to move. This process was repeated three times for each inclination angle on both SOCAL and capped SOCAL surfaces, with different positions selected for each trial. Prior to data extraction using pyDSA, the video recordings were binarized and rotated to facilitate the detection of the droplet contour when it was sliding. This pre-processing step was performed through the Camtasia software. A third-degree polynomial fitting was used to extract the time, as well as the droplet's position, base radius, front contact angle, and back contact angle. The videos were analysed at intervals of every 10 frames, rather than every frame, as the percentage error between data points remained below 0.5%. This interval was therefore chosen to improve efficiency without comprising the precision of the measurements. For each droplet experiment at a given substrate tilt angle, the position-time graphs were analyzed to determine the droplet had reached a steady speed.

## 3. Molecular-Kinetic Theory (MKT) type Model for Droplet with Circular Contact Area

*Modified MKT* – From Blake and Haynes [R7], the local interface velocity in the direction perpendicular to the contact line is given by,

$$v = 2\lambda K_0 \sinh\left[\frac{W}{k_B T}\right] \tag{S3.1}$$

In this equation, $W = W(\varphi)$ is the net work done to move a given point of the contact line (determined by the azimuthal angle $\varphi$), $k_B$ is the Boltzmann constant, $T$ is the temperature, $\lambda$ is the average distance between adsorption sites on the surface and the molecular displacement occurs with an average frequency $K_o$. For a moving droplet we assume that $W(\varphi)$ corresponds to the net work done in displacing the interface from an average dynamic contact angle $\theta_d$ to the local dynamic angle $\theta(\phi)$, i.e.,

$$W(\varphi) = \lambda^2 \gamma_{LV}(\cos\theta_d - \cos\theta(\varphi)), \tag{S3.2}$$

where we assume, for simplicity, that

$$\cos\theta_d = \frac{1}{2}(\cos\theta_f + \cos\theta_b) \tag{S3.3}$$

We are interested in the average speed of the droplet in the direction of motion, $v_x = v\hat{n} \cdot \hat{e}_x$, where $\hat{n} = (\cos\varphi, \sin\varphi)$ is the unit outward normal to the contact line and $\hat{e}_x = (1,0)$ is the unit vector in the $x$-direction. Hence, we obtain:

$$v_x(\varphi) = 2\lambda K_0 \sinh\left[\frac{W(\varphi)}{k_B T}\right] \cos\varphi \tag{S3.4}$$

and the average speed is then,

$$v_T = \frac{1}{2\pi r_c} \int v_x(\varphi) r_c d\varphi \tag{S3.5}$$

The integral cannot be evaluated in terms of elementary functions to yield an explicit expression for the average speed. However, it can be written as a Fourier cosine series as,

$$\cos\varphi \sinh[\alpha(\cos\theta_d - \cos\theta(\varphi))] = a_o(1 + a_1\cos\varphi + a_2\cos 2\varphi + \cdots)\sinh\left[\frac{\alpha(\cos\theta_b - \cos\theta_f)}{2}\right] \tag{S3.6}$$

where the coefficients $a_n$ are $a_0$=0.5, $a_2$=1 and all other $a_n$=0 for $\alpha \leq 0.5$ for all $\theta_f$ and $\theta_b$. For $\alpha > 0.5$, the series captures the periodicity, $a_0$ decreases in value and is no longer independent of $\theta_f$ and $\theta_b$, $a_2 \neq 1$ and other terms are non-zero. In performing the average in eq. S3.5, all terms other than $a_0$ average to zero, i.e.

$$\frac{1}{2\pi}\int_0^{2\pi} \cos\varphi \sinh[\alpha(\cos\theta_d - \cos\theta(\varphi))] = a_o \sinh\left[\frac{\alpha(\cos\theta_b - \cos\theta_f)}{2}\right] \tag{S3.7}$$

Figure S1 shows the Fourier series approximation (dashed line) to the original equation (solid line) in eq. 3.6 for $\theta_f$= 180° and $\theta_b$=0° for (a) $\alpha$=0.5 and (b) (a) $\alpha$=5. The Fourier series for Fig. S1b includes terms up to $a_{20}$ and the first term $a_o$=0.328.

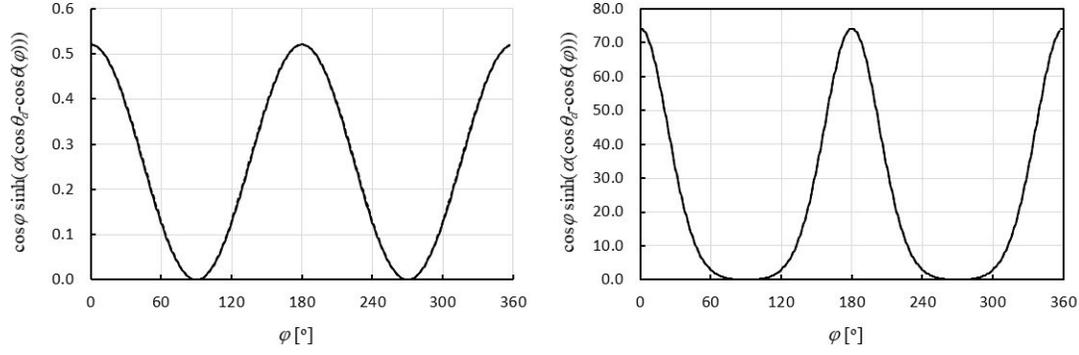

FIG. S1 shows the Fourier series approximation (dashed line) compared to the original equation (solid line) in eq. S3.6 for $\theta_f = 180°$ and $\theta_b = 0°$ for (a) $\alpha = 0.5$ and (b) (a) $\alpha = 5$. The dashed line and solid line cannot easily be visibly distinguished.

When $\alpha \leq 0.5$ the average droplet speed is given by:

$$v_T \approx \lambda K_0 \sinh\left[\frac{\lambda^2 \gamma_{LV}(\cos\theta_b - \cos\theta_f)}{2k_B T}\right] \tag{S3.8}$$

or, in terms of the driving force $F_f = 2k\gamma_{LV} r_c (\cos\theta_b - \cos\theta_f)$:

$$v_T \approx v_S \sinh\left[\frac{\lambda^2 F_f}{4kr_c k_B T}\right] \tag{S3.9}$$

where $v_S = \lambda K_0$.

*Validity of the Approximation* – We note that,

$$\alpha[\cos\theta_d - \cos\theta(\varphi)] = \frac{1}{2}\alpha \cos\varphi \left(\cos\theta_b - \cos\theta_f\right) \tag{S3.10}$$

and, in our case we have small differences between the front and back contact angles, we find,

$$\alpha[\cos\theta_d - \cos\theta(\varphi)] \approx \frac{1}{2}\alpha \cos\varphi \, \Delta\theta \sin\theta_{ave} \tag{S3.11}$$

The range of validity $a_o = 0.5$ is when the argument of sinh() is equal to unity, i.e.

$$\alpha \cos\varphi \, \Delta\theta \sin\theta_{ave} \leq 2 \tag{S3.12}$$

Or using the maximum in $\cos\varphi$,

$$\alpha \leq \frac{2}{(\cos\theta_b - \cos\theta_f)} \approx \frac{2}{\Delta\theta \sin\theta_{ave}} \tag{S3.13}$$

In physical parameters this gives,

$$\frac{\lambda^2 \gamma_{LV}}{k_B T} \leq \frac{2}{(\cos\theta_b - \cos\theta_f)} \approx \frac{2}{\Delta\theta \sin\theta_{ave}} \tag{S3.14}$$

$$\lambda \leq \sqrt{\frac{2k_BT}{\gamma_{LV}(\cos\theta_b - \cos\theta_f)}} \approx \sqrt{\frac{2k_BT}{\gamma_{LV}\Delta\theta \sin\theta_{ave}}} \tag{S3.15}$$

For water with a surface tension $\gamma_{LV}=0.0728$ N/m at T=293 K, this gives,

$$\lambda \leq 3.33 \times 10^{-10}\sqrt{\frac{1}{(\cos\theta_b - \cos\theta_f)}} \approx 3.33 \times 10^{-10}\sqrt{\frac{1}{\Delta\theta \sin\theta_{ave}}} \tag{S3.16}$$

For our surfaces, $\sin\theta_{ave} \approx 1$ and taking $\Delta\theta$ in degrees,

$$\lambda \leq 3.33 \times 10^{-10}\sqrt{\frac{180}{\pi\Delta\theta}} = 2.53 \times 10^{-9}\sqrt{\frac{1}{\Delta\theta}} \tag{S3.17}$$

Thus, for front and back contact angle differences of $\Delta\theta \geq 1°$ the approximation will be valid for fits with $\lambda < 2.5$ nm.

*Linearized Response* – Expanding Eq. (S3.9) gives a velocity-frictional force relationship for a moving droplet,

$$v_T \approx \frac{v_s \lambda^2 F_f}{4kr_c k_B T} \tag{S3.18}$$

The frictional force response with changes of velocity is therefore,

$$\frac{dF_f}{dv_T} \approx \frac{\pi r_c k_B T}{v_s \lambda^2} = \frac{\pi r_c k_B T}{K_0 \lambda^3} \tag{S3.19}$$

This can be compared to the frictional force-velocity response defined using a friction coefficient, $\beta$, defined in Li *et al.* [R8],

$$\frac{dF_f}{dv_T} = 2\beta r_c \eta \tag{S3.20}$$

Equating the two equations gives

$$\beta = \frac{\pi k_B T}{2\eta K_0 \lambda^3} \tag{S3.21}$$

and

$$v_s = \frac{\pi k_B T}{2\eta \beta \lambda^2} \tag{S3.22}$$